# Interface-engineered voltage-driven magnetic tunnel junctions with ultra-low-energy magnetization switching

Yu Zhang[1*], Meng Xu[1*], Bowei Zhou[1], Carter Eckel[1], Supriya Ghosh[2], Hwanhui Yun[2,6], Ali Habiboglu[1], Deyuan Lyu[3], Daniel B Gopman[4], Jian-Ping Wang[3],

K. Andre Mkhoyan[2] and Weigang Wang[1,5] ✉

1. Department of Physics, University of Arizona, Tucson, Arizona 85721, USA
2. Department of Chemical Engineering and Materials Science, University of Minnesota, Minneapolis, Minnesota 55455, USA
3. Department of Electrical & Computer Engineering, University of Minnesota, Minneapolis, MN 55455, USA
4. Materials Science & Engineering Division, National Institute of Standards and Technology, Gaithersburg, MD 20899, USA
5. Department of Electrical & Computer Engineering, University of Arizona, Tucson, Arizona 85721, USA
6. Korea Research Institute of Chemical Technology, Dajeon 34114, South Korea

*Contributed equally

✉ Email: wgwang@arizona.edu

**Electric-field control of spin states offers a promising route to ultra-low-power, ultra-fast magnetization switching in spintronic devices such as magnetic tunnel junctions (MTJs). Recent progress in modulating spin–orbit interactions at the interfaces between 3*d* transition-metal ferromagnets and dielectric layers has underscored the role of atomic-scale heavy-metal doping in optimizing device performance. Here, we experimentally demonstrate highly energy-efficient, voltage-driven magnetization switching in MTJs exhibiting large tunnel magnetoresistance (TMR), enabled by a remote doping technique that precisely controls the iridium (Ir) concentration near the MgO/CoFeB interface in the free layer. Our devices achieve a switching energy of only 3.5 fJ per bit for nanoscale MTJs operating in the sub-nanosecond regime, while maintaining a TMR ratio up to 160 % after 400 °C post-annealing. These findings establish a viable pathway toward scalable, ultra-low-power nonvolatile memory, positioning voltage-driven MTJs as strong contenders for next-generation magnetoresistive random-access memory (MRAM) and other emerging spintronic applications.**

Magnetoresistance random-access memory (MRAM) has numerous advantages over dynamic random-access memory (DRAM) such as low power consumption, fast access speed and unlimited endurance[1–3]. The magnetic tunnel junction (MTJ)[4–8] is the key component of MRAM cells, with the core structure comprising two ferromagnetic (FM) layers (i.e., a free layer and a reference layer) separated by a thin tunnel barrier. The data bits are stored as the magnetization configuration of ferromagnetic layers, either parallel (P) as "0" or anti-parallel (AP) as "1", leveraging tunneling magnetoresistance to yield lower resistance for the "0" state. A significant breakthrough in MRAM was achieved with the advent of the spin-transfer torque (STT) effect, which allows the magnetization of a ferromagnet to be switched with a spin-polarized current instead of a magnetic field[9,10]. An important feature for STT-MRAM is that the write current and read current paths identically pass through the tunnel barrier (i.e., MgO), which could lead to the undesired reliability issues, including the read disturbance and MgO degradation, particularly at switching speed below 10 ns[11]. An alternative method to switch the free layer is through spin-orbit torque (SOT) effect[12–18]. Distinct from STT-MRAM, the spin current in SOT-MRAM is generated by the spin-orbit coupling effects in a non-magnetic layer adjacent to the free layer. In this fashion, the writing-current path becomes separated from the reading-current path in a three-terminal device structure, which naturally avoids the writing-reading interference and MgO degradation. However, due to the orthogonal configuration of the spin polarization of spin current (e.g., *y* direction) and the equilibrium magnetization direction (e.g., *z* direction of perpendicular MTJs), the SOT does not directly compete with the Gilbert damping torque, resulting an even higher switching current density in SOT-MRAMs. The lowest switching energy obtained so far through current-induced switching mechanisms is 45 fJ, achieved in double spin-torque MTJs[19]. In recent years, the voltage-controlled magnetic anisotropy (VCMA) effect has been utilized as one of the promising writing principles to achieve energy efficient switching in MRAMs[20–42]. For 3*d*-transition-metal ferromagnets, itinerant *d* electrons with energy close to the Fermi level generate the magnetic moment and dictate magnetic anisotropy energy through orbital occupation within the *d* band, and consequentially the magnetic anisotropy energy $K_u$ can be modified through modifying the orbital state occupation with an external electric field. This kind of VCMA effect can take place at the interface of ferromagnetic material and dielectric material (e.g., Fe-MgO or CoFe-MgO), induced by either selective electron-hole doping in *d*-electron orbitals[21,24,32,36] or electric-field induced magnetic dipole moment[23]. Since the VCMA effect has purely electronic origins and is free from the ionic migration[43–45], this phenomenon permits ultra-fast voltage-driven magnetization switching of MTJ by using sub-nanosecond voltage pulses. Moreover, since no large current is needed in the scheme of voltage-driven magnetization switching process, energy efficient magnetization manipulation can be accomplished due to the substantially reduced Joule heating. A small write energy down to 6 fJ was recently reported for Ta-CoFeB-MgO based VCMA MTJs with tunnel magnetoresistance (TMR) ratio up to 45%[41,42]. Further reduction of writing energy needs lower switching voltages, which demands a larger VCMA effect. First principle calculations suggest that a larger VCMA effect can be enabled by introducing heavy metal atoms at and/or near the interface of ferromagnetic layer and MgO layer[22,32]. Indeed it was demonstrated that a large VCMA coefficient of 320 fJV$^{-1}$m$^{-1}$ can be achieved in Iridium (Ir)-doped Fe/MgO junctions thanks to the enhanced modulation of the 5*d* majority spin states of the interfacial and sub-interfacial Ir atoms [25,27]. However, it is well known that the introduction of impurity atoms at the barrier interface is detrimental to coherent tunneling, making it highly advantageous to preserve a pristine FM/MgO interface[46,47]. More importantly, the potential improvement in energy efficiency from this approach has not yet been validated in MTJ nanodevices, as dynamic switching in the sub-nanosecond regime remains largely unexplored.

In this work, we introduce a remote doping technique in CoFeB/MgO junctions that enhances VCMA efficiency without degrading the TMR, thereby significantly improving device performance. Our approach builds on two key observations in VCMA-based MTJs: while spin-dependent tunneling is a strictly interfacial phenomenon—dominated by contributions from the first and second FM layers adjacent to the tunnel barrier[48]—the VCMA effect is more "non-local", as suggested by previous studies[25,47,49]. Accordingly, remote Ir doping away from the MgO interface, followed by controlled diffusion during post-growth thermal annealing[50], can lead to an optimal Ir distribution that enhances the VCMA effect while preserving coherent electron tunneling across the CoFeB/MgO interface. By modulating the Ir doping profile, we not only achieve high TMR and strong perpendicular magnetic anisotropy (PMA), but also optimize the switching probability of voltage-driven precessional switching in MTJ nanopillars. Our results demonstrate that ultra-low writing energy of 3.5 fJ can be achieved in the sub-nanosecond regime, while maintaining a TMR ratio up to 160 % after 400 °C post-annealing.

The schematic structures of four types of MTJ prepared in this study are shown in **Figure 1a**. These MTJ films include the buffer layers of Ta (5)/Ru(10)/Ta(5) and the caping layers of Ta(7.5)/Ru(15), where the numbers in parentheses indicate the layer thicknesses in nanometers. The core stack structure of the four MTJs are: Mo(1)/CoFeB(0.85)/MgO(2.5)/CoFeB(1.5)/Mo(1.5) (Sample 1), Mo(1)/CoFeB(0.85)/MgO(2.5)/Ir(0.05/0.1/0.15)/CoFeB(1.5)/Mo(1.5) (Sample 2), Mo(1)/CoFeB(0.85)/MgO(2.5)/CoFeB(1.5)/Ir(0.1)/Mo(1.5) (Sample 3) and Mo(1)/CoFeB(0.85)/MgO(2.5)/CoFeB(0.75)/Ir(0.1)/CoFeB(0.75)/Mo(1.5) (Sample 4). The bottom CoFeB serves as the reference ferromagnetic layer in these junctions. After deposition, the MTJ films were then patterned into nanodevices of 160 nm in diameter by optical lithography, a nanosphere lithography (NSL) technique[51] and ion beam etching[52,53]. Then the patterned MTJ nanodevices were annealed at different temperatures from 300 °C to 500 °C in an Ar atmosphere. Additional information regarding sample fabrication can be found in our prior work[54–56]. **Figure 1b** schematically illustrates the microwave circuit used for evaluating the TMR ratio and switching probability of voltage-driven magnetization switching. More details of the measurement setup and device characterization can be found in the **Methods**.

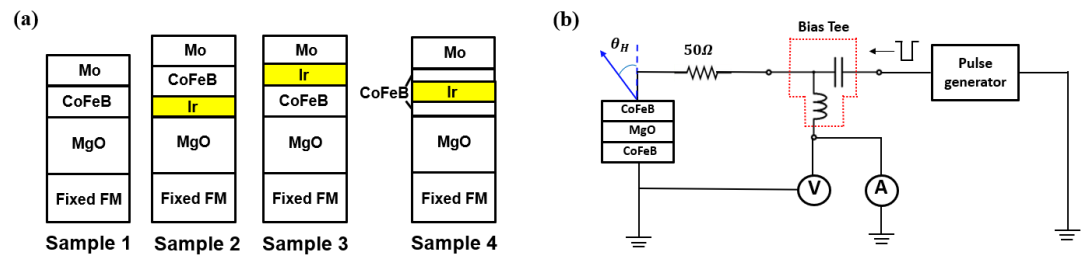

**Figure 1 | (a) MTJ stacks for Sample 1, 2, 3 and 4. A thin Ir layer was inserted into MTJ free layer with different position for Sample 2, 3 and 4. (b) A schematic of measurement setup.**

We first studied the influence of Ir insertion on TMR ratio and PMA of MTJs. **Figure 2a** shows the TMR curves of MTJs with different locations of Ir insertion layer. These devices were annealed at 300 °C. Notably there is no obvious TMR difference between Sample 1 (no Ir insertion) and Sample 3 (0.1 nm Ir layer inserted at the CoFeB/Mo interface), but the TMR decreases more than 40 % when 0.1 nm Ir layer was inserted inside top CoFeB (Sample 4). **Figure 2b** shows the TMR curves of MTJs with different thickness of Ir insertion layer at the interface of MgO and top CoFeB (Sample 2). The TMR keeps decreasing as the thickness of Ir insertion layer increases from 0 nm to 0.15 nm. The results of both **Figure 2a** and **2b** were then summarized as 3D plot in **Figure 2c.** On one hand, the TMR shows a clear decrease with the increase of inserted Ir thickness at MgO/CoFeB interface (Sample 1, 2), which is a direct result of the destruction of coherent tunnelling channels[57]. On the other hand, TMR shows an obvious increase when the location of Ir insertion layer gradually gets far away from the MgO/CoFeB interface while keeping the Ir thickness at 0.1 nm (Sample 3, 4). In a previous study it was found that both VCMA and TMR can be increased when a CoFe layer is introduced between the Ir-doped FM electrode and the MgO barrier[47]. Our Sample 4 has a similar structure but it doesn't exhibit high TMR as shown in **Figure 1a** and **1c**. Therefore, in the subsequent discussion we will focus on the other three samples.

An annealing temperature of 400 °C is required for the compatibility with back end of line (BEOL) process of CMOS technology[58]. Therefore, the influence of Ir insertion on Coercivity ($H_c$) (**Figure 2d**) and TMR ratio (**Figure 2e**) were also investigated under different annealing temperatures from 300 °C to 500 °C. In the subsequent discussion, only 0.1 nm Ir insertion was used for Sample 2, as it was the smallest thickness we can fabricate with high reproducibility. As shown in **Figure 2d**, the $H_c$ of all three samples reduces with the increasing of annealing temperature, indicating PMA deteriorates due to the diffusion of heavy metal atoms into MgO/CoFeB interface[59]. For the relative lower annealing temperatures at 300 °C and 350 °C, the $H_c$ for Sample 1 and Sample 3 are nearly same, which are almost double for $H_c$ of Sample 2. This result indicates that the PMA of Sample 2 can be degraded even at a lower annealing temperature if Fe-O and/or Co-O bonds at MgO/CoFeB interface is interrupt by Ir insertion[60,61]. In fact, the $H_c$ of Sample 2 is consistently smaller than the other two samples under all annealing conditions. These findings suggest the PMA of Sample 3 is more thermally robust than that of Sample 2 at 400 °C, which is in a consistence with previous study[54]. It is clear from these results that although the insertion of Ir generally has a negative influence on PMA, the impact can be much reduced if the Ir is placed at the CoFeB/Mo interface rather than the MgO/CoFeB interface.

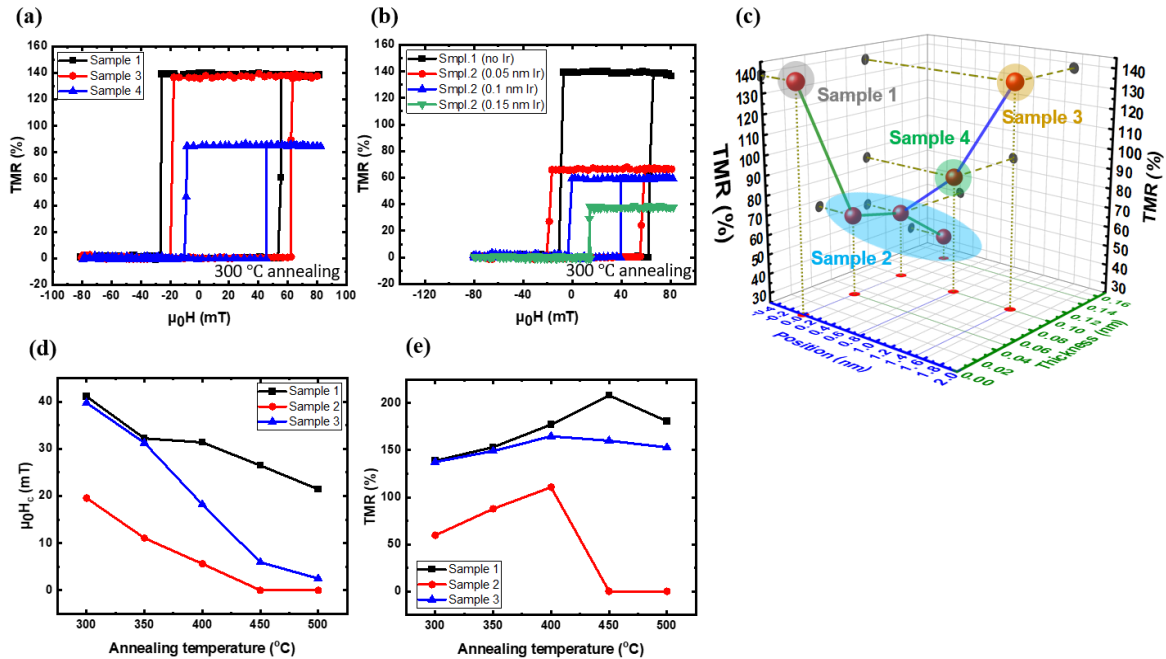

**Figure 2** | (a) Curves of TMR-H for MTJs with different positions of Ir insertion layer. (b) Curves of TMR-H for Sample 2 with different thickness of Ir insertion layer. The black curve is for Sample 1 (without Ir insertion layer). (c) TMR evolves with different thickness and different position of Ir insertion layer. The data points of Sample 1, Sample 2, Sample 3 and Sample 4 are marked as the background color of grey, blue, yellow and green, respectively. (d) Coercivity and (e) TMR ratio for Sample 1, 2 and 3 under different annealing temperatures. Only 0.1 nm Ir insertion was used for Sample 2.

The evolution of TMR with different annealing temperatures is summarized in **Figure 2e**. As the annealing temperature increasing from 300 °C to 400 °C, all three samples show sizable increases in TMR, which can be explained by the further crystallization of CoFeB at higher temperatures[7,62–65]. However, compared with Sample 1 and Sample 3, Sample 2 shows a lower TMR ratio under all annealing temperatures. Furthermore, when the annealing temperature increases to 450 °C and 500 °C, the Sample 2 quickly loses its TMR owing to the loss of PMA[64], which has been indicated from **Figure 2d**. At the same time, the TMR of Sample 1 (Sample 3) reaches a value of 208 % (165 %) at 450 °C (400 °C). Note these TMR ratios are generally much higher than those obtained in nanopillars in previous VCMA dynamic switching studies[41,42]. The decrease of TMR in Sample 1 (Sample 3) after annealing at 500 °C (450 °C) is likely related to the diffusion of Mo/Ta (Ir) atoms[54] into the upper CoFeB and MgO layers. These results suggest that the negative influence of Ir to TMR can be largely mitigated by remotely doping Ir at the CoFeB/Mo interface (Sample 3), likely due to the reduced Ir doping density near MgO/CoFeB interface, especially at the annealing temperature of 400 °C and above.

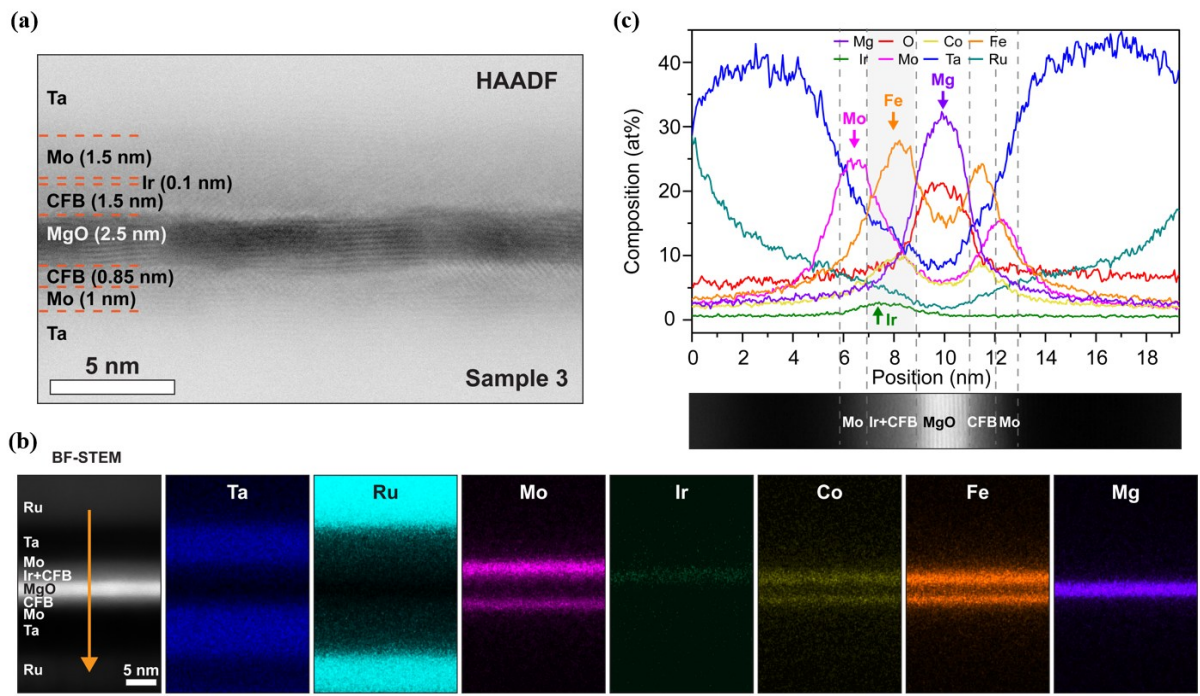

**Figure 3 | (a)** HAADF-STEM image of Sample 3 cross-section with post-annealing of 400 °C. **(b)** BF-STEM image and corresponding EDX elemental maps for Ta, Ru, Mo, Ir, Co, Fe and Mg in the core structure of Mo(1 nm)/CoFeB(0.85 nm)/MgO(2.5 nm)/CoFeB(1.5 nm)/Ir(0.1 nm)/Mo(1.5 nm) in Sample 3. **(c)** Elemental composition scan profile across the selected section in BF-STEM imaging. The peaks of the signal for Mo, Ir and Fe and Mg elements are marked with arrows for eye-guide.

In order to investigate the difference of Ir doping density near MgO/CoFeB interface between Sample 2 and Sample 3, microstructural characterization and elemental composition analyses were performed using scanning transmission electron microscopy (STEM). **Figure 3a** shows the high-angle annular dark-field (HAADF)-STEM image from the device cross-section of Sample 3 after 400 °C post-annealing. A highly oriented polycrystalline MgO (001) layer can be observed, which serves as the template to crystallize the CoFe(B) into a bcc structure associated with high TMR[62,66]. The bright-field (BF)-STEM imaging and corresponding energy dispersive X-ray spectroscopy (EDX) maps are shown in **Figure 3b**. A clear edge between Mo and Ir layers is observed, while the Ir layer partially overlaps with the CoFe (B) layer. This result is further demonstrated in the elemental composition analysis profile shown in **Figure 3c**, where the peak of the signal for elements Mo, Ir and Fe are spatially separated with each other. In more details, the distance between the peaks of Ir and Mg is around 2.91 nm and the Ir elemental concentration decreases by 52% from its peak position to the MgO/CoFeB interface. As a comparison, the **Supplementary Information-1** shows the EDX mapping and elemental composition analysis for Sample 2. From the elemental composition analysis profile, the distance between the peaks of Ir and Mg is obvious narrower for Sample 2, which is around 1.46 nm; and the Ir elemental concentration only decreases by 33 % from its peak position to the MgO/CoFeB interface. Those results clearly demonstrate that a lower Ir doping density near MgO/CoFeB interface is present in Sample 3, which has been introduced by remotely inserting Ir layer at the CoFeB/Mo interface. Another interesting observation is that the peak of Ir elements is found to be moved towards CoFe(B) layer, in both Sample 2 and Sample 3. We attribute this phenomenon to the kinetics of thin film growth and annealing between different crystallinity. The sputtered CoFeB thin films is amorphous as deposit, while both Mo (i.e., CoFeB/Ir/Mo in Sample 3) and MgO (i.e., MgO/Ir/CoFeB in Sample 2) layers are crystalline. The diffusion of Ir atoms is preferrable towards the metastable amorphous structure, i.e., CoFeB layer during the film deposition and post-annealing.

Next, we study the influence of inserted Ir layer on the switching probabilities of voltage-driven magnetization switching for MTJs under different annealing temperatures. A product of switching probabilities ($P_{sw}$) defines as $P_{sw}=P_{01}*P_{10}$, where $P_{01}$ ($P_{10}$) is the single switching probability from P (AP) to AP (P) configuration. **Figure 4a, 4b and 4c** show the evolution of the $P_{sw}$ with increasing pulse voltages under different annealing temperatures. The applied voltages here are relatively small to prevent the MTJ nanodevices from breakdown. For the annealing temperature at 300 °C, $P_{sw}$ in all three samples firstly remains very low when pulse voltage ($V$) is below a threshold pulse voltage ($V_{th}$), and gradually increases to a higher level when $V$ exceeds $V_{th}$. In more details, when $V<V_{th}$, the VCMA effect is not sufficiently lowering down the energy barrier of MTJ. When $V=V_{th}$, the magnetization starts to process around the effective magnetic field $H_{eff}$, which is the sum of the following three components: magnetic anisotropy field under VCMA effect, the stray field from reference layer of MTJ and the external magnetic field. A smaller PMA and a larger VCMA coefficient result in a lower $V_{th}$. Based on the previous results in **Figure 2d**, Sample 2 has smallest PMA while Sample 1 has the largest PMA. Also consider the fact that VCMA coefficient can be largely enhanced in Ir-doped (Co)Fe/MgO junctions[25,27], that is why Sample 2 shows the lowest $V_{th}$ and then Sample 3, while Sample 1 shows the highest $V_{th}$. Sample 2 with a higher Ir doping density near MgO/CoFeB interface also shows a larger $P_{sw}$ at high pulse voltages, which is due to its larger VCMA coefficient than Sample 1 and Sample 3 with 300 °C post-annealing. These results are consistent with previous observations[22,25,27,29].

After annealing at 400 °C **(Figure 4b)**, while Sample 1 shows a similar behavior of $P_{sw}$ as at 300 °C, Sample 2 and Sample 3 exhibit totally different features. On the one hand, for Sample 2, $P_{sw}$ remains at nearly zero even under higher pulse voltages, because of the collapse of PMA due to increased Ir diffusing as shown in Figure 2d; on the other hand, a substantial increase of $P_{sw}$ is observed for Sample 3, which can be explained by a more optimal doping density of Ir near MgO/CoFeB interface under a higher annealing temperature. In more details, the higher annealing temperature of 400 °C enables an appropriate amount of Ir atoms to diffuse towards the MgO/CoFeB interface, forming a doping gradient of Ir element inside MTJ structure and a relatively low (but apparently sufficient) Ir doping density near MgO/CoFeB interface, as indicated in **Figure 3c.** Comparing to Sample 2, the relatively low Ir doping density near MgO/CoFeB interface can bring two obvious benefits: firstly, it helps to avoid the severe $P_{sw}$ degradation due to the loss of PMA at higher annealing temperatures; secondly and surprisingly, it can still induce a large VCMA coefficient, which contributes to a very high $P_{sw}$ at higher pulse voltages (as shown in **Figure 4b** for Sample 3). Therefore, by remotely doping Ir at the CoFeB/Mo interface in Sample 3, the desired properties of high TMR and high PMA of Sample 1, as well as the high VCMA coefficient of Sample 2, can be obtained at the same time.

As for the annealing temperature of 500 °C, a further increased Ir doping density near MgO/CoFeB interface led to a rapid decrease of PMA in Sample 3, resulting to a zero $P_{sw}$, as shown in **Figure 4c**. The $P_{sw}$ of Sample 1 also decreases substantially, due to the deterioration of the MTJ structure after annealing under such as high temperature. **Figure 4d** shows a summary of the $P_{sw}$ of three samples under different annealing temperatures. Sample 3 shows the highest $P_{sw}$ at 400 °C among all three samples, together with its high TMR ratio = 165 % (**Figure 2e**). Appropriately controlled Ir doping density near MgO/CoFeB interface combines the advantageous functional properties of high VCMA coefficient and high annealing temperature stability of both PMA and TMR, which is critical for low writing energy and high switching probability.

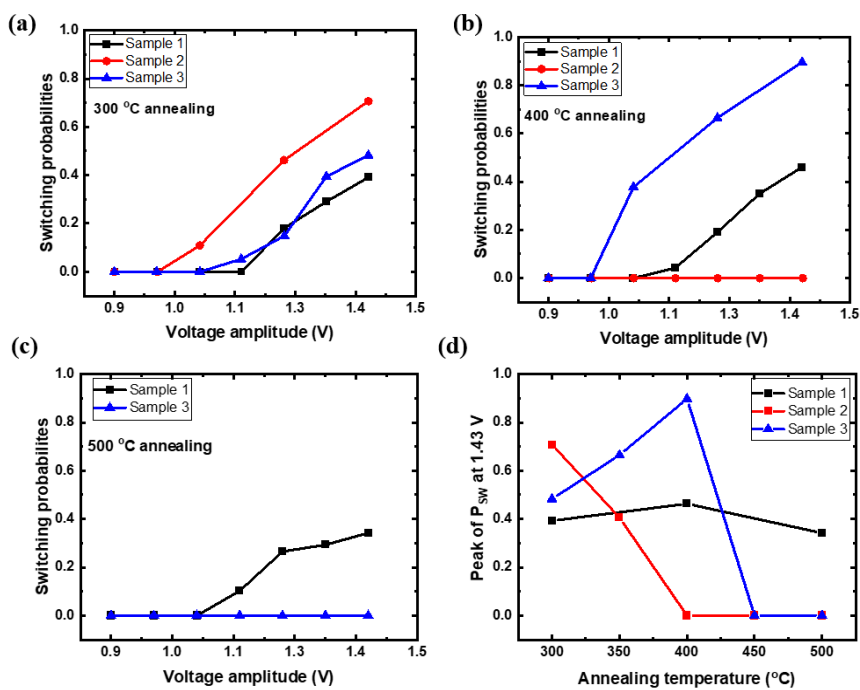

**Figure 4 | The product of switching probabilities as a function of pulse voltage under the annealing temperature of (a) 300 °C, (b) 400 °C and (c) 500 °C. (d) The peak of the product of switching probabilities for Sample 1, 2 and 3 under different annealing temperatures. The data of (d) comes from (a), (b) and (c).**

Finally, the detailed switching dynamics of the devices are presented in Figure 5. All the nanopillars shown here were annealed under the temperature of 400 °C. **Figure 5a** shows the measurements of resistance R versus the magnetic field H for Sample 1 at two different DC bias voltages: 10 mV and 600 mV (see the definition in **Figure 1b** and **Methods**). The $H_c$ of Sample 1 decreases dramatically to nearly zero when the applied DC voltage changes to a higher voltage. This behavior is well-known as VCMA effect[20,21,37,39]. Note the center of the minor TMR loop also shift dramatically under high voltage, due to the voltage-controlled interlayer coupling effect[67]. When voltage pulses with sufficient magnitude are applied, the MTJs can be switched by the mechanism as discussed before, where the switching probability oscillates with pulse width, as shown in **Figure 5b**. An example of successful back-and-forth precessional magnetization switching is shown in the inset, measured by the change of resistance after each identical voltage pulse application of 1.72 V. The high and low resistance states appear alternately, indicating the reversible magnetization switching of free layer happens with each successive voltage pulse. More detailed discussion for VCMA effect and the voltage-driven precessional magnetic switching can be found in **Supplementary Information-2**. As shown in **Figure 5b**, a $P_{sw}$ above 60 % was achieved with a pulse duration ≈ 100 ps and a magnitude of 1.72 V for Sample 1, showing the great potential for ultra-fast voltage-driven switching in the ≈ 100 ps scale. Here we point out that the further improvement of such switching speed for voltage-driven MTJs will be limited by PMA. More detailed discussion can be found in **Supplementary Information-3**.

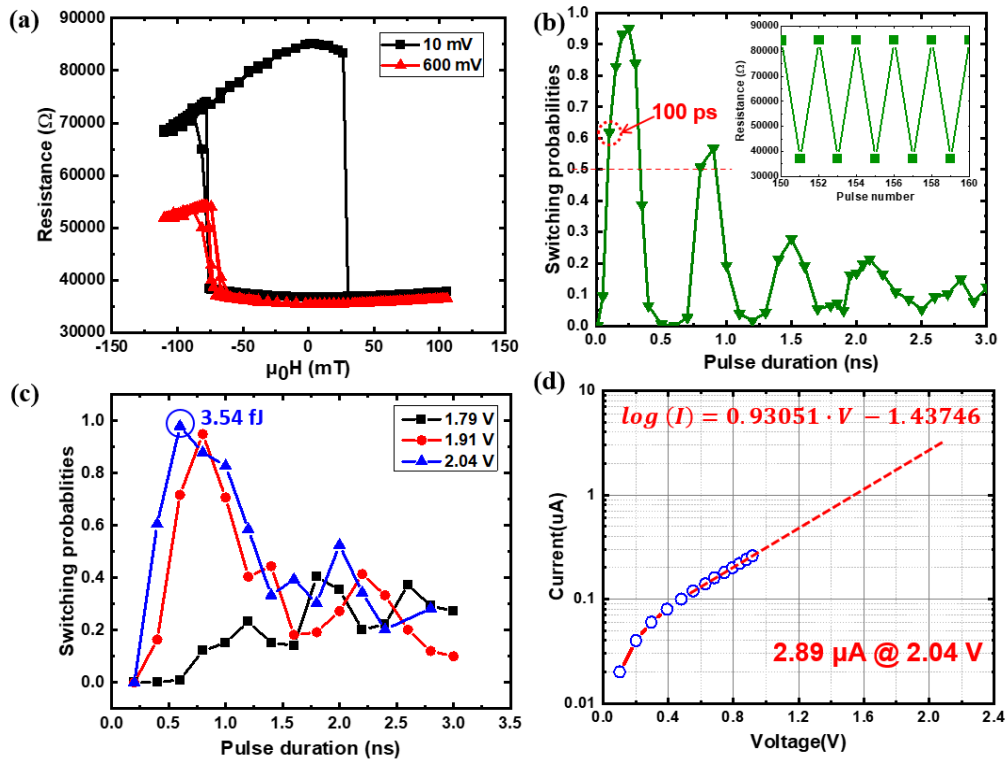

**Figure 5 | (a)** R-H loop of Sample 1. The $\theta_H$ of the applied magnetic field is fixed at 70 degrees. The large (600 mV) and small (10 mV) volage were applied to the MTJ and the resistance was measured by sweeping the magnetic fields. **(b)** The $P_{sw}$ as a function of pulse duration under a fixed voltage pulse of 1.72 V. The applied magnetic field is $\mu_0 H$ = -63.0 mT and the $\theta_H$ of the applied magnetic field is fixed at 70 degrees. A magnetic switching with a $P_{sw}$ = 62% was achieved with 100 ps voltage pulse duration. The inset shows the toggle switching behavior with the pulse number between 150 and 160. The voltage pulses are fixed at 1.72 V with 0.9 ns duration, and the magnetic magnitude and the $\theta_H$ of the magnetic field are $\mu_0 H$ = -25.8 mT and 40 degrees. **(c)** The $P_{sw}$ as a function of pulse duration with different voltage magnitudes under a fixed magnetic field of $\mu_0 H$ = -28.0 mT and the $\theta_H$ of the applied magnetic field is fixed at 55 degrees. As marked with blue solid circle, a magnetic switching with a $P_{sw}$ = 98% was achieved with the writing energy as low as 3.54 fJ/bit, voltage pulse magnitude ~ 2.04 V and pulse duration ~ 0.6 ns. **(d)** I-V curve for Sample 3, which shows how to estimate the switching current through linear fitting of current by voltage.

The dependence of switching probability on pulse duration of Sample 3 is presented in **Figure 5c**. As commonly observed in dynamic voltage-control precessional switching, $P_{sw}$ of Sample 3 oscillates with the increase of pulse duration, and a higher $P_{sw}$ is achieved with the higher voltage pulse magnitude. A $P_{sw}$ close to 100 % can be achieved by a voltage pulse with a magnitude of 2.04 V with a pulse duration of 0.6 ns. The thermal stability ($\Delta = E_b/k_B T$, where $E_b$ is the anisotropy energy barrier, $k_B$ is the Boltzmann constant, and T is the temperature) of this device, estimated by the dependence of $H_c$ on the magnetic field ramp rate, is ~32, which is similar to the values (~ 23 to 30) obtained in MTJ nanopillars in previous VCMA dynamic switching studies[34,38,41]. To calculate the actual switching energy, the MTJ resistance (or the corresponding current) at 2.04 V is required, since the tunnelling resistance decreases dramatically at higher voltages. This is estimated from the *I-V* curve of the device using a method previously reported[42], as shown in **Figure 5d**. The switching energy is determined to be 3.54 fJ at the peak $P_{sw}$ value of Sample 3, which to the best of our knowledge is the lowest value obtained in MTJs. Note that the dimension of our MTJ nanodevice in this work is around 160 nm, limited by the nanosphere lithography in our fabrication process. The switching energy can in principle be further reduced by shrinking the size of MTJ nanodevice. The extrapolated writing energy for a 10 nm MTJ with optimally controlled Ir doping density near the MgO/CoFeB interface can reach approximately 0.01 fJ/bit, which is comparable in magnitude to the switching energy of typical CMOS transistor, showing the strong potential of the voltage-induced scheme reported here for ultra-low energy consumption.

In conclusion, by explicitly leveraging the interfacial nature of spin-dependent tunnelling and the non-local character of VCMA, we demonstrated a low switching energy of 3.5 fJ for nanoscale MTJs operating in the sub-nanosecond regime, while maintaining a TMR ratio up to 160 % after 400 °C post-annealing. Through the investigation of PMA, TMR, switching probabilities, and microstructural characterization of nanodevices with various stack configurations, we found that an optimal Ir doping density near the MgO/CoFeB interface can be effectively introduced via remote doping of Ir at the CoFeB/Mo interface. This discovery is critical for achieving a high VCMA coefficient while mitigating degradation of PMA and TMR. These energy-efficient, voltage-driven MTJ nanodevices with ultra-fast switching speeds offer strong potential for the next generation MRAM and other spintronics applications.

**Methods**

**Sample preparation.**
All samples in this study were deposited onto silicon wafers with 1 μm of thermal oxide layer using a combination of direct current and radio frequency magnetron sputtering, in a commercial sputter deposition system with a base pressure of ~$10^{-9}$ Torr. The substrates were kept at ambient temperate during deposition. The metallic layers were deposited using DC magnetron sputtering sources, while MgO layer was deposited using an RF magnetron sputtering source.

**Device nanofabrication and rapid-annealing.**
The deposited films were patterned into 160 nm MTJ nanodevices using a combination of conventional optical lithography, ion beam etching (IBE), and nanosphere lithography (NSL). Details of the NSL technique can be found in Reference[51]. The patterned samples were conducted a rapid-annealing treatment in Ar atmosphere under different temperatures ranging from 300 °C to 500 °C to optimize the PMA and TMR ratio.

**Sample characterization.**
The magnetic properties of the post-annealed thin films were characterized using vibrating sample magnetometry. The coercivity ($H_c$) was characterized from the measured hysteresis loops.

**Device characterization.**

The fabricated MTJ nanodevices were connected into the microwave measurement setup using *rf* probes, as shown in the **Figure 1b**. Sub-nanosecond voltage pulses with variable amplitudes and duration were generated by a pulse generator and applied to the devices through the *rf* port of a bias tee. The actual pulse voltages applied on the MTJ nanodevices are almost doubled due to the impedance mismatch between MTJ and the 50 Ω coaxial tip and transmission line[41]. The DC port of the bias tee was connected to a DC current source and a voltmeter to monitor the resistance of MTJ nanodevices to calculate the TMR ratio. The TMR ratio is defined as $(R_{AP}-R_P)/R_P$, where $R_P$ and $R_{AP}$ are the MTJ resistance in P and AP configuration, respectively. Tilted external magnetic fields with varying magnitudes and polar angles $\theta_H$ were applied to MTJ nanodevices during the measurements. The magnetization switching probability was determined by measuring the resistance of MTJ nanodevices 500 times after applying identical voltage pulses. All the measurements were performed at room temperature.

**Structural characterization and elemental compositional analysis.**

Cross-sectional STEM samples were prepared using a dual-beam focused ion beam (FIB) system. A protective layer consisting of amorphous carbon and platinum was deposited onto the sample surface to prevent ion beam damage during high-current milling steps involved in lamella lift-out. The samples were thinned using a 30 keV $Ga^+$ ion beam, followed by low-energy milling at 2 keV to remove surface damage layers. HAADF-STEM imaging and EDX mapping were conducted using an aberration-corrected FEI (S)TEM microscope, equipped with a monochromator, CEOS-DCOR probe corrector, and an EDX spectrometer. The microscope was operated at 200 keV, with a probe convergence angle of 18.2 mrad and HAADF detector collection angles between 55 and 200 mrad. Elemental maps and HAADF images were acquired using a probe current of 100 pA.

**Acknowledgements**

The authors thank Dr. Roland Himmelhuber for technical support in nanofabrication. The authors also thank Dr. Jiaqi Zhou for fruitful discussions. This work was primarily supported in part by the Defense Advanced Research Projects Agency (DARPA) under Grant No. HR001117S0056-FP-042 and the National Science Foundation (NSF) under Grant No. ECCS- 2230124. The work was also partially supported by NSF through award No. DMR-2309431. STEM work was carried out in the Characterization Facility, University of Minnesota, which receives partial support from the NSF through the MRSEC (DMR-2011401) and the NNCI (award number ECCS2025124) programs. S.G. acknowledges support from a Doctoral Dissertation Fellowship received from the Graduate School at the University of Minnesota. H.Y. would like to acknowledge the institutional research program (KK2452-20) and support program for young researchers (BSK24-131) funded by KRICT.

**Author contributions**

Y.Z., M.X. and W.W. conceived and designed the devices. Y.Z., M.X., B.Z, A.H., D.L. and J.W. carried out the material deposition and device fabrication. Y.Z., M.X., B.Z., C.E., D.B.G. and D.L. performed the measurements. S.G., H.Y. and K.A.M. conducted the structural and compositional characterization. Y.Z. and W.W. wrote the manuscript with input from all authors. All authors discussed the results, contributed to the data analysis, and reviewed the manuscript. The study was performed under the supervision of W.W.

**Competing financial interests**
The authors declare no competing financial interests.

# Supplementary Information

**Supplementary Information-1**
The **Figure (a)** below shows the HAADF-STEM image of Sample 2 after 300 °C post-annealing. The selected region used for EDX mapping and elemental composition analysis is marked with red rectangle in Figure (a), which has the core structure of Mo(1 nm)/CoFeB(0.85 nm)/MgO(2.5 nm)/Ir(0.1 nm)/CoFeB(1.5 nm)/Mo(1.5 nm). As shown in **Figure (b)**, EDX elemental mapping of Ir, Co, Fe, Mg, and O elements clearly reveals the diffusion of Ir atoms into the CoFe(B) layer and toward the MgO/CoFeB interface. This observation is further supported by the elemental composition analysis profile in **Figure (c)**. As shown in **Figure (c)**, the peak of element Ir is closer to the peaks of element Mg and O, comparing to the corresponding profile shown in **Figure 3c** in the main text. This indicates a higher Ir doping density near the interface of MgO/CoFeB in the free layer in Sample 2.

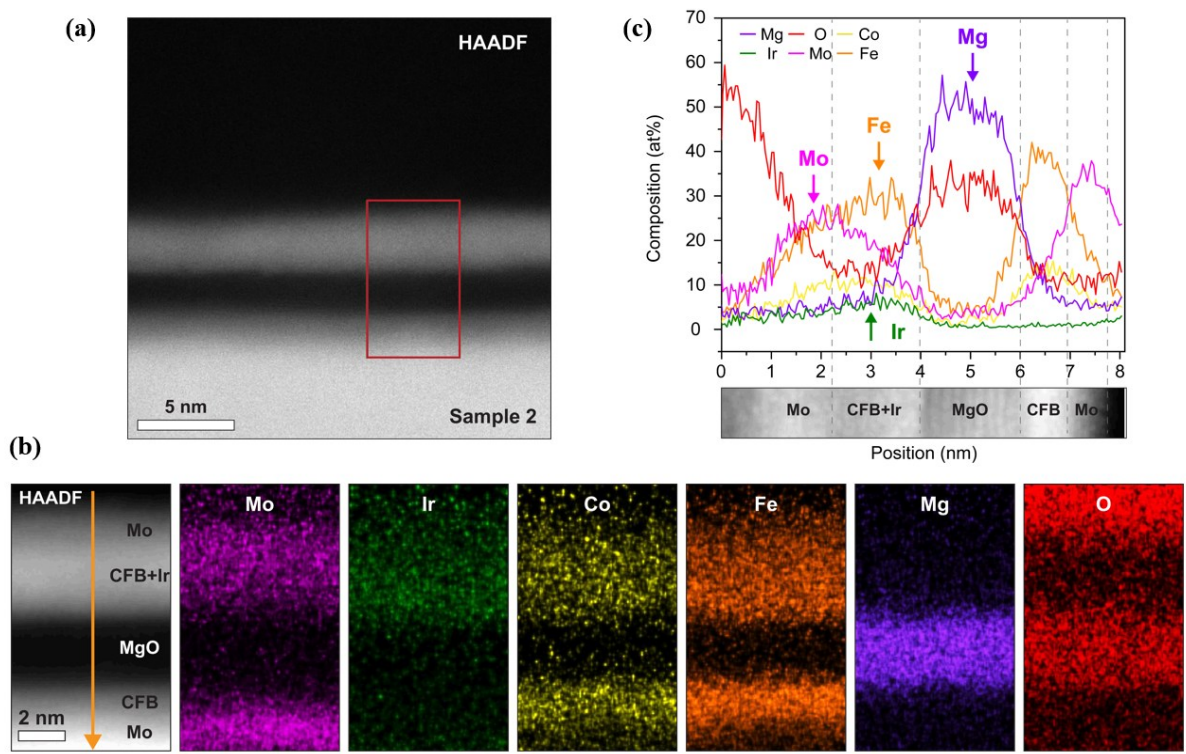

**Supplementary Information-2**

When the PMA is nearly reduced to zero by a short voltage pulse (see **Figure 5a**) applied on MTJs, the magnetization of free layer CoFeB starts to precess around the in-plane component of the external magnetic field. The precession frequency is determined by the amplitude of this in-plane field. If the voltage pulse stops at the half of a precession period, the magnetization can be switched to the opposite direction and then stabilized through relaxing. Therefore, the switching time $t_{sw}$ equals to the half of the procession period and can be further expressed as

$$t_{sw} \sim \frac{\pi(1-\alpha^2)}{\gamma\mu_0 H_{in-plane}} \qquad \text{[\textbf{Equation 1}]}$$

Where α, γ and $\mu_0$ are the Gilbert damping factor, the gyromagnetic ratio and the permeability of vacuum, respectively[29][Reference: Micromachines 10, (2019)].

The table below shows the experimental data of Sample 1. The high switching probability can be only achieved when the magnetic field and pulse duration satisfy certain conditions. For the magnetic field, on one hand, since the perpendicular component need to compensate the stray field from reference layer, for magnetic fields with different polar angles $\theta_H$, their perpendicular components are nearly equal to each other (the data marked with the light-yellow color background); and on the other hand, the in-plane component of magnetic field is used to provide the axis of precessional switching (the data marked with light-green color background). From **Figure (a)**, a linear relationship between the switching time and the oscillation period can be observed. The linear fitting gives a slope around 2, which can be understood that when the pulse duration equals to half turn precession, the fully switching can be achieved. From **Figure (b)**, there is another linear relationship between the switching speed and the in-plane component of magnetic field. Actually, this linear dependence can be described by the equation as inset, which indicates the nature of the Larmor precession: Here, the $H_{eff}$ is the in-plane component of magnetic field and $\tau$ is the period of precession. Therefore, we can see that, when the stray field from the reference layer is fully compensated, the switching speed is only determined by in-plane component of magnetic field.

The precessional dynamics of the MTJ magnetization are reflected in the following two aspects: firstly, the oscillation of switching probability ($P_{sw}$) as a function of pulse width, as shown in the **Figure 5b** (Sample 1) and **5c** (Sample 3). Secondly, the toggle switching behavior for the resistance of MTJs with successive voltage pulses of the same voltage magnitude and pulse duration, as shown in the inset of **Figure 5b** (Sample 1).

| External magnetic field | | | Pulse | | Oscillation period ($\tau$) | Switching probability |
| --- | --- | --- | --- | --- | --- | --- |
| Angle (degree) | Magnitude ($\mu_0 H$ (mT)) | | Voltage (V) | Duration (ns) | | |
| | Total | In-plane | perpendicular | | | |
| 40 | −25.8 | −16.6 | −19.8 | 1.72 | 0.9 | 1.7 | 99% |
| 50 | −30.1 | −23.1 | −19.3 | | 0.6 | 1.4 | 96% |
| 60 | −44.2 | −38.3 | −22.1 | | 0.4 | 0.8 | 100% |
| 70 | −63.0 | −59.2 | −21.5 | | 0.2 | 0.5 | 97% |

$$H_{eff} = H_{ext} \sin \theta_H \qquad H_{stray} \approx H_{ext} \cos \theta_H$$

**Table 1**

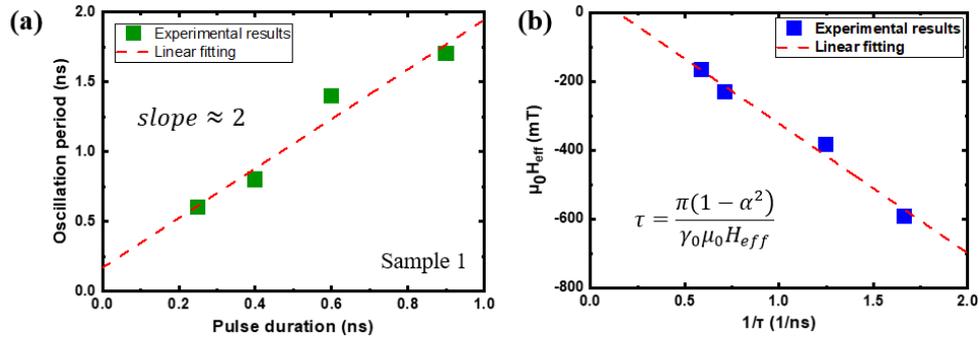

(a) slope ≈ 2, Sample 1

(b) $\tau = \dfrac{\pi(1-\alpha^2)}{\gamma_0 \mu_0 H_{eff}}$

**Supplementary Information-3**

More experimental data of Sample 1 are shown in the **Figure (a)** to **(d)** below. **Figure (a)** shows the how to find the optimized magnetic field from the curves of switching probabilities for each fixed polar angle $\theta_H$. The optimized external magnetic field can be determined by scanning the magnetic field to get the maximum of the product of switching probabilities under each polar angle of magnetic fields. **Figure (b)** shows the oscillatory behavior in precessional switching under a fixed external magnetic field of $\mu_0 H = -25.8$ mT and $\theta_H = 40$ degrees. The maximum of the product of switching probabilities increases with the increasing of voltage pulse magnitude from 1.53 V to 1.72 V.

In **Figure (c)**, a series of voltage pulses with the fixed magnitude ~ 1.72 V were applied to the MTJ nanodevices. The $\theta_H$ of external magnetic field changed from 40, 50, 60 and 70 degrees, while the magnitude of field changed from $\mu_0 H = -25.8$ mT, −30.1 mT, −44.2 mT to −63.0 mT, respectively. Note that the perpendicular component of external magnetic field keeps constant to compensate the stray field from reference layer in MTJ, while the varied in-plane component serves as the in-plane axis for Larmor precession of magnetization. More details have been discussed in **Supplementary Information-2**. With the increase of in-plane component of magnetic field, the oscillating curves shift towards left, which is in a consistence with **Equation 1** [see **Supplementary Information-2**]. **Figure (d)** shows the zoomed-in feature of **Figure (c)** for pulse duration below 1 ns. As shown in **Figure (d)** (Sample 1), a $P_{sw}$ above 60% was achieved with a pulse duration ~ 100 ps, showing the great potential for ultra-fast voltage-driven switching in the ~100 ps scale. We would like to point out that further improvement of such voltage-driven magnetic switching speed can be limited by PMA. As previously shown in **Equation 1**, the voltage-driven magnetic switching speed is proportional to the in-plane component of external magnetic field ($H_{in\text{-}plane}$). Therefore, a faster precessional switching can be facilitated by a larger $H_{in\text{-}plane}$. However, for perpendicular magnetized-MTJs, the increase of $H_{in\text{-}plane}$ will lead to the decrease of thermal stability by lowering down the energy barrier between P and AP states given by $\Delta K_B T$, as following[26,40][Reference: Appl. Phys. Express 9, (2016); Phys. Rev. Appl. 10, 024004 (2018)]:

$$\Delta = \Delta_0 \left(1 - \frac{H_{in-plane}}{H_k}\right)^2 \quad \text{[Equation 2]}$$

Where $H_k$ is the anisotropy field, and $\Delta$ and $\Delta_0$ ($\equiv \mu_0 H_k M_s v / 2k_B T$, where $M_s$ and $v$ denote the saturation and volume of the free layer, respectively) are the thermal stability factors with and without $H_{in\text{-}plane}$, respectively. Since the decrease of thermal stability will lead to the instability of information storage, therefore a large enough PMA is essential for MTJs to keep two distinct magnetization states (i.e., P and AP configurations) when an in-plane field is applied. As a result, the PMA becomes the limitation for applying larger in-plane fields for an even faster precessional switching. The coercivity values in **Figure 2d** indicate that Sample 1 has the largest PMA among all three samples under 400 °C annealing, that is why the fastest switching speed as 150 ps is only achieved in Sample 1 in our study, as shown in **Figure (c)** and **(d)**.

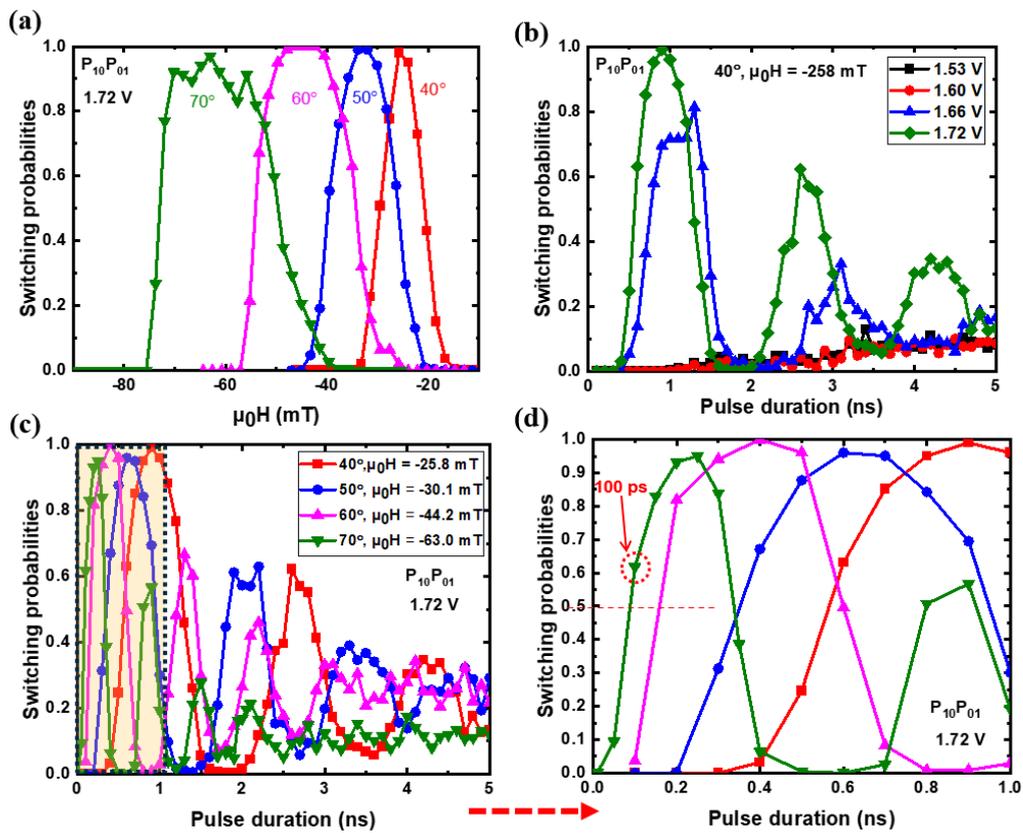